\documentclass[aps,pra,showpacs,superscriptaddress,preprint]{revtex4}
\usepackage{graphicx}

\catcode`ð=\active

 \defð{\u{g}}

 \catcode`Ð=\active

 \defÐ{\u{G}}

 \catcode`Ý=\active

 \defÝ{\. I}

 \catcode`ö=\active

 \defö{\"{o}}

 \catcode`Ö=\active

 \defÖ{\"O}

 \catcode`ü=\active

 \defü{\"{u}}

 \catcode`Ü=\active

 \defÜ{\"{U}}

 \catcode`Þ=\active

 \defÞ{\c{S}}

 \catcode`þ=\active

 \defþ{\c{s}}

 \catcode`ý=\active

 \defý{{\i}}

 \catcode`ç=\active

 \defç{\d{c}}

 \catcode`Ç=\active

 \defÇ{\d{C}}

\begin{document}
\title{Exact Solutions of Effective Mass Dirac Equation
with non-$PT$-Symmetric and non-Hermitian Exponential-type
Potentials}
\author{\small Altuð Arda}
\email[E-mail: ]{arda@hacettepe.edu.tr}\affiliation{Department of
Physics Education, Hacettepe University, 06800, Ankara,Turkey}
\author{\small Ramazan Sever}
\email[E-mail: ]{sever@metu.edu.tr}\affiliation{Department of
Physics, Middle East Technical  University, 06800, Ankara,Turkey}

\date{\today}

\begin{abstract}
By using two-component approach to the one-dimensional effective
mass Dirac equation bound states are investigated under the effect
of two new non-$PT$-symmetric, and non-Hermitian, exponential type
potentials. It is observed that the Dirac equation can be mapped
into a Schr\"{o}dinger-like equation by rescaling one of the two
Dirac wave functions in the case of the position dependent mass.
The energy levels, and the corresponding Dirac eigenfunctions are found analytically.\\
Keywords: Dirac equation, Position-Dependent Mass,
non-$PT$-symmetric potential
\end{abstract}
\pacs{03.65.-w; 03.65.Ge; 12.39.Fd}

\maketitle

\newpage
The investigation of the quantum systems having so-called
non-Hermitian Hamiltonians has been a great interest because of
its theoretical contributions to quantum mechanics [1-15]. The
study of such quantum systems has been received many applications,
especially in quantum field theories [16-18], and nuclear theory
[19]. The solutions of the non-relativistic and relativistic
equations with non-Hermitian Hamiltonians having real or complex
energy spectra have also been studied by many authors by using
different methods [3, 6, 11, 20]. Non-Hermitian Hamiltonians
satisfy the condition given by
$\hat{O}H\hat{O}^{-1}=\hat{O}H\hat{O}=H$, where the operator
$\hat{O}\equiv PT$ is an operator combined with parity, and
time-reversal transformations, respectively. So they are related
with $PT$-symmetry [8]. If a potential $V(x)$ has the condition
written as $V(-x)=V^{\star}(x)$ under the transformation of $x
\rightarrow -x$, and $i \rightarrow -i$, then it is said that the
potential is a $PT$-symmetric.

The relativistic wave equations, especially Dirac equation, give
numerous important results, and explanations in the view of
quantum mechanics. One of them is the spin-orbit coupling, and
related topics, such as the Hall effect, spin torque,
Zitterbewegung, etc [21-23]. Many authors have also been studied
the solution of the Dirac equation with different potentials in
different theoretical backgrounds [24, 25].

Recently, the generalization of the solutions of the relativistic
and non-relativistic wave equations with $PT$-symmetric potentials
in the case of the constant mass to the case of the
position-dependent mass become very attractive research topic.
Many authors have studied the effects of the position-dependent
mass on the solutions of the above equations [26-36]. The new
formalism based on the position-dependent mass is a useful ground
to study the properties of some physical systems, such as quantum
dots [37], semiconductor heterostructures [38], quantum liquids
[39].

In this letter, we intend to study the effects of the
position-dependent mass on the solutions of the one-dimensional
Dirac equation by using the map between the mass, and potential
functions without the spin-effects [43]. Our aim, by taking two
examples of the mass distributions, is to show that the
two-component approach [40, 41] to the one-dimensional Dirac
equation can be used in the position-dependent mass formalism. We
obtain two new non-$PT$-symmetric, and non-Hermitian potential
functions produced by using the map between the mass, and
potential functions.

The relativistic Dirac equation in the absence of an external
potential can be written as
\begin{eqnarray}
\bigg(i\gamma^{\mu}\frac{\partial}{\partial
x^{\mu}}-m\bigg)\Psi=0\,,
\end{eqnarray}
where $m$ is the rest mass of the particle, and $\gamma^{\mu}\,
(\mu=0,1,2,3)$ are gamma matrices ($\hbar=c=1$). The Dirac
equation for a particle moving in an external potential $V(x)$ in
one-dimension can be written as [40]
\begin{eqnarray}
[\alpha\,.\,p+\beta m(x)-(E-V(x))]\,\psi(x)=0\,,
\end{eqnarray}
where $E$ is the relativistic energy of the particle, $p$ is the
momentum operator, $m(x)$ is the mass function and $\alpha, \beta$
are $2 \times 2$ matrices which are set to Pauli matrices
$\sigma_3$ and $\sigma_1$, respectively. The Dirac wave function,
$\psi(x)$, is decomposed into an upper $\psi^+ (x)$, and lower
component $\psi^- (x)$, so that
\begin{eqnarray}
\psi(x)=\Bigg(\begin{array}{c}
\psi^{+}(x) \\
\psi^{-}(x)
\end{array}\Bigg)\,,
\end{eqnarray}

We have the following set of two coupled differential equations by
inserting Eq. (3) into Eq. (2)
\begin{eqnarray}
\frac{d\psi^{+}(x)}{dx}\,+\,[m(x)+(E-V(x))]\psi^{-}(x)=0\,,\\
\frac{d\psi^{-}(x)}{dx}\,+\,[m(x)-(E-V(x))]\psi^{+}(x)=0\,.
\end{eqnarray}

We define the wave functions $f(x)$ and $g(x)$ in two-component
approach which makes possible to obtain the solutions easily
without the spin effects [40, 41] as
\begin{eqnarray}
\Bigg(\begin{array}{c}
f(x) \\
g(x)
\end{array}\Bigg)=\Bigg(\begin{array}{cc}
1 & i \\
1 & -i
\end{array}\Bigg)\,\Bigg(\begin{array}{c}
\psi^{+}(x) \\
\psi^{-}(x)
\end{array}\Bigg)\,,
\end{eqnarray}
and insert into Eqs. (4) and (5), we have
\begin{eqnarray}
\frac{df(x)}{dx}\,-\,i[E-V(x)]f(x)+im(x)g(x)=0\,,\\
\frac{dg(x)}{dx}\,+\,i[E-V(x)]g(x)-im(x)f(x)=0\,.
\end{eqnarray}

By using the transformation written in terms of mass function
$m(x)$ in Eq. (7)
\begin{eqnarray}
f(x)=\sqrt{m(x)\,}\,\phi(x)\,,
\end{eqnarray}
one gets
\begin{eqnarray}
i\,\frac{d\phi(x)}{dx}\,+\,\Bigg[\,\frac{im'(x)}{2m(x)}\,+(E-V(x))\Bigg]\phi(x)-\,\sqrt{m(x)\,}g(x)=0\,.
\end{eqnarray}
where prime denotes the derivative to the spatial coordinate. To
obtain a Schr\"{o}dinger-like equation for $\phi(x)$, one sets the
potential term $V(x)$ in Eq. (10) as
\begin{eqnarray}
V(x)=i\frac{m'(x)}{2m(x)}\,,
\end{eqnarray}
which gives
\begin{eqnarray}
\frac{d^2\phi(x)}{dx^2}\,-V_{eff}(x)\phi(x)=-E^2\phi(x)\,.
\end{eqnarray}

This is the Schr\"{o}dinger equation having the energy eigenvalue
$E^2$ under the effect of the potential $V_{eff}(x)=m^2(x)$. By
choosing a suitable mass function, one can construct a potential
$V(x)$ from Eq. (11). By solving the Schr\"{o}dinger-like equation
given by Eq. (12), one obtains the energy spectra and
corresponding spinors of the Dirac equation. The Schr\"{o}dinger
and Schr\"{o}dinger-like equations can be solved by using the
calculation procedure, which suggests a hypergemetric type
equation in the following form [42]
\begin{eqnarray}
\frac{d^{2}\psi(z)}{dz^{2}}+\frac{\tilde{\tau}(z)}{\sigma(z)}
\frac{d\psi(z)}{dz}+\frac{\tilde{\sigma}(z)}{\sigma^{2}(z)}\psi(z)=0\,,
\end{eqnarray}
where $\sigma(z)$, $\tilde{\sigma}(z)$ are polynomials at most
second degree and $\tilde{\tau}(z)$ is a first degree polynomial.

By using the transformation $\psi(z)=\chi(z)y(z)$, one gets
\begin{eqnarray}
\sigma(z)\,\frac{d^2
y(z)}{dz^2}\,+\,\tau(z)\,\frac{dy(z)}{dz}\,+\lambda y(z)=0\,,
\end{eqnarray}

The first part of the total wave function $\chi(z)$ is defined as
[42]
\begin{eqnarray}
\frac{\chi^{\prime}(z)}{\chi(z)}=\frac{\pi(z)}{\sigma(z)}\,,
\end{eqnarray}
and the other part of the solution $y(z)$ is given by the
Rodrigues relation
\begin{eqnarray}
y_{n}(z)=\frac{a_{n}}{\rho(z)}\frac{d^{n}}{dz^{n}}
\left[\sigma^{n}(z)~\rho(z)\right],
\end{eqnarray}
where $a_n$ is a normalization constant, and $\rho(z)$ is the
weight function obtained from the following relation [42]
\begin{eqnarray}
\frac{d}{dz}\left[\sigma(z)~\rho(z)\right]=\tau(z)~\rho(z).
\end{eqnarray}

The polynomial $\pi(z)$ in Eq. (15), and the parameter $\lambda$
required for the method are given by
\begin{eqnarray}
\pi(z)=\frac {\sigma^{\prime}(z)-\tilde{\tau}(z)}{2}\pm
\sqrt{\left(\frac{\sigma^{\prime}(z)-\tilde{\tau}(z)}{2}\right)^{2}-
\tilde{\sigma}(z)+k\sigma(z)\,}\,.
\end{eqnarray}
\begin{eqnarray}
\lambda=k+\pi^{\prime}(z)\,.
\end{eqnarray}
The discriminant of the expression under the square root in the
polynomial $\pi(z)$ in Eq. (18) must be zero, which defines the
constant $k$. Thus, a new eigenvalue equation becomes
\begin{eqnarray}
\lambda=\lambda_{n}=-n\tau'(z)-\frac{n(n-1)}{2}\,\sigma''(z)\,.
\end{eqnarray}
where $\tau(z)=\tilde{\tau}(z)+2\pi(z)$ has a negative derivative.
The energy eigenvalues are obtained from the Eqs. (19) and (20).

We prefer the following mass function as a first case
\begin{eqnarray}
m(x)=m_0(1+e^{-\,\delta x})\,,
\end{eqnarray}
which gives an exponential type, non-$PT$-symmetric, non-Hermitian
potential
\begin{eqnarray}
V(x)=\,-\frac{i\delta}{2}\frac{e^{-\,\delta x}}{1+e^{-\,\delta
x}}\,.
\end{eqnarray}
By using the mass function, Eq. (12) gives
\begin{eqnarray}
\frac{d^2\phi(x)}{dx^2}\,+\,\Big\{E^2-m^2_0-2m^2_0e^{-\,\delta
x}-m^2_0e^{-2\,\delta x}\Big\}\phi(x)=0\,,
\end{eqnarray}

By inserting the new variable $z^{-1}=e^{\delta x}$, one obtains
the following equation
\begin{eqnarray}
\frac{d^2\phi(z)}{dz^2}\,+\,\frac{1}{z}\,\frac{d\phi(z)}{dz}\,+\,\frac{1}{z^2}\,
\Big\{\tilde{E}^2-\alpha^2-2\alpha^2 z-\alpha^2
z^2\Big\}\phi(z)=0\,.
\end{eqnarray}
where $\tilde{E}^2=E^2/\delta^2$, and $\alpha^2=m^2_0/\delta^2$.
Comparing Eqs. (24) and Eq. (13), and using the parameters
$-a_1=-\alpha^2$, $-a_2=-2\alpha^2$, and
$-\epsilon=\tilde{E}^2-\alpha^2$, we obtain
\begin{eqnarray}
\tilde{\tau}(z)=1\,\,\,;\,\,\,\,\,\sigma(z)=z\,\,\,;\,\,\,\,\,\tilde{\sigma}(z)=-a_1
z^2-a_2 z-\epsilon\,.
\end{eqnarray}

To obtain the polynomial $\pi(z)$, one inserts these polynomials
into Eq. (18), with $\sigma\,'(z)=1$, then we get
\begin{eqnarray}
\pi(z)=\pm \sqrt{a_1 z^2+(k+a_2)z+\epsilon\,}\,,
\end{eqnarray}
We obtain two solutions for $k$ as $k_{1,2}=-a_2 \pm2\sqrt{a_1
\epsilon\,}$, from the condition that the discriminant of the
expression under the square root has to be zero. Eq. (18) gives
\begin{displaymath}
\pi(z)=\left\{ \begin{array} {ll} \pm
[\,\sqrt{a_1\,}z+\sqrt{\epsilon\,}\,], & k \rightarrow k_{1}\,,\\
\pm [\,\sqrt{a_1\,}z-\sqrt{\epsilon\,}\,], & k \rightarrow
k_{2}\,.
\end{array} \right.
\end{displaymath}
We find the polynomial $\tau(z)$ by choosing $k_2$ as
\begin{eqnarray}
\tau(z)=\tilde{\tau}(z)+2\pi(z)=1+2\sqrt{\epsilon\,}-2\sqrt{a_1\,}z\,,
\end{eqnarray}
which gives $\tau\,'(z)=-2\sqrt{a_1\,}<0$. Thus, an energy
eigenvalue equation is obtained from Eqs. (19) and (20) as
\begin{eqnarray}
(2n+1)\sqrt{a_1\,}+2\sqrt{a_1 \epsilon\,}+a_2=0\,,
\end{eqnarray}
\begin{eqnarray}
E=\pm\sqrt{\,m^2_0-\,\frac{\delta^2}{4}\,\Big[\,2n+1+\,\frac{2m_0}{\delta}\,\Big]^2\,}\,.
\end{eqnarray}

It can be seen that the energy levels of particle and antiparticle
are symmetric about zero. We observe that it is obtained zero
energy in the case of the limit $\delta \rightarrow 0$\, which is
the energy level of the following equation derived from Eq. (23)
\begin{eqnarray}
\frac{d^2\phi(x)}{dx^2}+\omega^2_{0}\phi(x)=0\,,
\end{eqnarray}
where $\omega^2_{0}=E^2-m^2_{0}$\,. The Dirac Hamiltonian with a
general scalar potential, $V'(x)$, gives always zero-energy
solutions in the ultrarelativistic limit [43], where the upper and
lower components given in Eq. (2) are written as
\begin{eqnarray}
\psi^{+}(x)\sim e^{-(mx+h(x))}\,,\\
\psi^{-}(x)\sim e^{+(mx+h(x))}\,.
\end{eqnarray}
with $h(x)=\int^{x}V'(y)dy$\,, $m$ denotes the rest mass of
particle. The normalization conditions for zero-energy eigenstates
are discussed in Ref. [43]. The solutions of Eq. (30) have the
same form with the ones given in Eqs. (31) and (32) for the limit
$\delta \rightarrow 0$\,, i.e., $h(x)=0$\, for $m \rightarrow
2m_{0}$\,. One has to find first the weight function $\rho(z)$ to
obtain the Dirac wave function $\phi(z)$. Using Eq. (17), we get
\begin{eqnarray}
\rho(z)=e^{-2\sqrt{a_1\,}z}\,z^{2\sqrt{\epsilon\,}}\,
\end{eqnarray}
and from Eq. (16), we obtain
\begin{eqnarray}
y_{n}(z)=\frac{a_{n}}{e^{-2\sqrt{a_1\,}z}\,z^{2\sqrt{\epsilon\,}}}\frac{d^{n}}{dz^{n}}
\left[e^{-2\sqrt{a_1\,}z}\,z^{n+2\sqrt{\epsilon\,}}\right].
\end{eqnarray}
From the last equation, we can write $y_n(z)$ in terms of the
generalized Laguerre polynomials as $y_n(z)\simeq
L^{(2\sqrt{\epsilon\,})}_{n}(z)$ [43]. The other part of the
solution is obtained from Eq. (15) as
\begin{eqnarray}
\chi(z)=e^{-\sqrt{a_1\,}z}\,z^{\sqrt{\epsilon\,}}\,.
\end{eqnarray}

Finally, we obtain the unnormalized wave function as
\begin{eqnarray}
\phi_{n}(z)=e^{-\sqrt{a_1\,}z}\,z^{\sqrt{\epsilon\,}}\,L^{(2\sqrt{\epsilon\,})}_{n}(z)\,.
\end{eqnarray}
Thus, we write the upper component from Eq. (9) as
\begin{eqnarray}
f_{n}(z)=\sqrt{m_0(1+z)\,}\,e^{-\sqrt{a_1\,}z}\,z^{\sqrt{\epsilon\,}}\,L^{(2\sqrt{\epsilon\,})}_{n}(z)\,
\end{eqnarray}
and the lower component from Eq. (8) as
\begin{eqnarray}
g_{n}(z)=\frac{e^{-\sqrt{a_1\,}z}\,z^{\sqrt{\epsilon\,}}\,L^{(2\sqrt{\epsilon\,})}_{n}(z)}
{\sqrt{m_0(1+z)\,}}\,\Big\{-i\delta
z\Big[\,\frac{\sqrt{\epsilon\,}}{z}\,-\sqrt{a_1\,}-\,\frac{L^{(2\sqrt{\epsilon\,})}_{n-1}(z)}
{L^{(2\sqrt{\epsilon\,})}_{n}(z)}\,+E\Big]\Big\}\,.
\end{eqnarray}

It is seen that the mass of the Dirac particle contributes to the
eigenfunctions, and the upper component depends on the energy of
the particle.

We choose the second mass distribution
\begin{eqnarray}
m(x)=\,\frac{m_0}{1+e^{-\,\delta x}}\,,
\end{eqnarray}
which gives following exponential type, non-$PT$-symmetric and
non-Hermitian potential from Eq. (11) as
\begin{eqnarray}
V(x)=\frac{i\delta}{2}\,\frac{1}{1+e^{\delta x}}\,.
\end{eqnarray}

It has the form of the Woods-Saxon potential for $i\delta/2
\rightarrow -V_0$ [45]. Substituting Eq. (39) into Eq. (12), and
using the new variable $z=1/(1+e^{-\,\delta x})$, we obtain
\begin{eqnarray}
\frac{d^2\phi(z)}{dz^2}\,+\,\frac{1-2z}{z(1-z)}\frac{d\phi(z)}{dz}-\,\Big\{\,\frac{\tilde{E}^2}{z^2(1-z)^2}\,
+\,\frac{\alpha^2}{(1-z)^2}\,\Big\}\phi(z)=0\,,
\end{eqnarray}
where $-\tilde{E}^2=E^2/\delta^2$. Comparing the Eqs. (41) and Eq.
(13), we get
\begin{eqnarray}
\tilde{\tau}(z)=1-2z\,\,;\,\,\,\,\sigma(z)=z(1-z)\,\,;\,\,\,\tilde{\sigma}(z)=-\tilde{E}^2-\alpha^2
z^2\,.
\end{eqnarray}

Following the same procedure as in last section, we get the energy
spectra
\begin{eqnarray}
E=\pm\,\frac{1}{4}\sqrt{8m^2_0-[\sqrt{\delta^2+4m^2_0\,}-\delta
(2n+1)]^2 -\Big(\,\frac{4m^2_0}{\sqrt{\delta^2+4m^2_0\,}-\delta
(2n+1)}\,\Big)^2\,}\,.
\end{eqnarray}
and the corresponding wave functions are written in terms of the
Jacobi polynomials
\begin{eqnarray}
\phi_n
(z)=z^{\tilde{E}}\,(1-z)^{-M}\,P^{(2\tilde{E}\,,\,-2M)}_{n}(1-2z)\,.
\end{eqnarray}

The lower and upper components $g_n (z)$ and $f_n (z)$ are written
by using Eqs. (44) and (8) respectively as
\begin{eqnarray}
f_{n}(z)=\sqrt{m_0\,}\,z^{\tilde{E}+1/2}\,(1-z)^{-M}\,P^{(2\tilde{E}\,,\,-2M)}_{n}(1-2z)\,,
\end{eqnarray}
and
\begin{eqnarray}
g_{n}(z)&=&z^{\tilde{E}-1/2}\,(1-z)^{-M}\,P^{(2\tilde{E}\,,\,-2M)}_{n}(1-2z)\,
\Big\{\,\frac{i\delta}{\sqrt{m_0\,}}\Bigg[\,\frac{1}{2}(1-z)+\tilde{E}(1-z)\nonumber\\&+&M
z-z(1-z)(n+1+2\tilde{E}-2M)\,\frac{P^{(1+2\tilde{E}\,,\,1-2M)}_{n-1}(1-2z)}{P^{(2\tilde{E}\,,\,-2M)}_{n}(1-2z)}\,\Bigg]
+\frac{E}{\sqrt{m_0\,}}\Big\}\,.
\end{eqnarray}
where $M=\sqrt{\tilde{E}^2+\alpha^2\,}$.

The energy levels of the particles and antiparticles are symmetric
about zero in Eq. (43). The non-$PT$-symmetric and non-Hermitian
potential given by Eq. (40) has a real energy spectrum under the
condition that $8m^2_0>[\sqrt{\delta^2+4m^2_0\,}-\delta
(2n+1)]^2+\Big(\,\frac{4m^2_0}{\sqrt{\delta^2+4m^2_0\,}-\delta
(2n+1)}\,\Big)^2$\,. The energy expression in Eq. (43) gives zero
energy in the limit of $\delta \rightarrow 0$ as in the first mass
distribution case. Both of the eigenfunctions are dependent on the
rest mass of the particle.

In summary, we have investigated the position-dependent mass
energy spectra, and the corresponding wave functions of the Dirac
equation in two-component approach. We have transformed it into a
Schr\"{o}dinger-like equation with the help of a transformation
applied on one of the Dirac wave functions by using two different
mass distributions. Thus two new non-$PT$-symmetric and
non-Hermitian complex potentials are produced. We have found that
the Dirac equation can be turned into a Schr\"{o}dinger-like
equation by using a suitable transformation depending on the mass
function $m(x)$ in two-component approach. This formalism gives us
two new non-$PT$-symmetric and non-Hermitian exponential and
inverse-exponential potentials. We have obtained a complex energy
spectrum for the exponentially potential given by Eq. (22), and
the potential given by Eq. (43) has a real energy levels if
$8m^2_0>[\sqrt{\delta^2+4m^2_0\,}-\delta
(2n+1)]^2+\Big(\,\frac{4m^2_0}{\sqrt{\delta^2+4m^2_0\,}-\delta
(2n+1)}\,\Big)^2$\,.

\end{document}